\documentclass{emulateapj}
\usepackage{threeparttable}
\usepackage{natbib}
\usepackage[bookmarks=false, colorlinks=true, citecolor=blue,
  backref=true]{hyperref}              
\hypersetup{
  colorlinks,
  citecolor=blue,
  linkcolor=red,
  urlcolor=blue,
}

\DeclareRobustCommand{\ion}[2]{%
\relax\ifmmode
\ifx\testbx\f@series
{\mathbf{#1\,\mathsc{#2}}}\else
{\mathrm{#1\,\mathsc{#2}}}\fi
\else\textup{#1\,{\mdseries\textsc{#2}}}%
\fi}

\shorttitle{}
\shortauthors{Cao et al.}

\begin{document}
\slugcomment{{\bf Accepted for publication in ApJ}}

\title{Ionized and cold gas components in low surface brightness galaxy AGC 102004}
\author{Tian-Wen Cao\altaffilmark{1}, Zi-Jian Li\altaffilmark{2,3,4}, Pei-Bin Chen\altaffilmark{1}, Chun-Yi Zhang\altaffilmark{1},
Gaspar Galaz\altaffilmark{5}, Cheng Cheng\altaffilmark{2,3}, Qingzheng Yu\altaffilmark{1},
Venu M. Kalari\altaffilmark{6}, Junfeng Wang\altaffilmark{1}, Hong Wu\altaffilmark{2}}
\altaffiltext{1}{Department of Astronomy, Xiamen University, 422 Siming South Road, Xiamen 361005, People$'$s Republic of China; \\astrocao@xmu.edu.cn, jfwang@xmu.edu.cn}
\altaffiltext{2}{National Astronomical Observatories, Chinese Academy of Sciences, Beijing 100101, People$'$s Republic of China;}
\altaffiltext{3}{Chinese Academy of Sciences South America Center for Astronomy, National Astronomical Observatories, Chinese Academy of Sciences, Beijing 100101, People$'$s Republic of China;}
\altaffiltext{4}{School of Astronomy and Space Sciences, University of Chinese Academy of Sciences, Beijing 100049, People$'$s Republic of China}
\altaffiltext{5}{Instituto de Astrofisica, Pontificia Universidad  Cat\'olica de Chile,\,$\!$\,Av.$\!$Vicu\~na Mackenna$\!$\,4860,\,7820436 Macul,$\!$\,Santiago, Chile}
\altaffiltext{6}{Gemini Observatory/NSFs NOIRLab, Casilla 603, La Serena, Chile}

\date{Accepted:25/07/2024}

\begin{abstract}

We present the integral field spectroscopic observations of ionized gas (H$\alpha$ and [{\ion{N}{II}}]) 
using the PCWI, along with deep CO(2-1) observations by the $^\backprime\bar{\rm U}^\backprime\bar{\rm u}$ receiver on JCMT
for AGC 102004. The velocity field of H$\alpha$ shows an anomalous distribution in the North-Western (NW) disk. 
The H$\alpha$ spectrum is well-fitted by two Gaussian components, 
and the weak Gaussian component is dominated by the anomalous H$\alpha$ in the NW disk.
The Gaussian fit center of H$\alpha$ emission is offset by +24.2 km s$^{-1}$ from the 
systemic velocity obtained from the HI emission.
We derive the gas-phase metallicity, 12+log(O/H), using [{\ion{N}{II}}]$\lambda$6583/H$\alpha$ ratio as a proxy.
The mean value of 12+log(O/H) is 8.30 $\pm$ 0.19 over the whole galaxy.
The metallicity in the outer disk is lower than the detection limit of 7.72, 
indicating the metallicity gradient exists in AGC 102004. 
We speculate a minor/mini-merger event could have happened to the NW disk.
CO(2-1) emission has non-detection in AGC 102004, reaching a noise level of 0.33 mK smoothed to 30 km s$^{-1}$.
The upper limit of molecular gas mass in AGC 102004 is 2.1 $\times$ 10$^7$ M$\odot$ with 
X$_{\rm CO}$ = 3.02$\times$10$^{20}$ cm$^{-2}$ (K km s$^{-1}$)$^{-1}$.
The M$_{\rm H_2}$/M$^{\rm corr}_{\rm HI}$ of AGC 102004 is lower than 0.0037 and lower than that of normal galaxies.

\end{abstract}


\section{Introduction \label{intro}}
Low surface brightness galaxies (LSBGs; \citealt{1987AJ.....94...23B, 
2000MNRAS.312..470B, 2011ApJ...728...74G, 2015AJ....149..199D, 2015ApJ...815L..29G, 2020ApJS..248...33H, 2021RAA....21...76C, 2021MNRAS.502.4262J, 
 2023A&A...676A..41J, 2023ApJ...953...83C, 2023ApJ...959..105D, 2023PASP..135j4101L, 2024RAA....24a5018Z, 2024ApJ...960....9Z, 2024ApJ...964...85D, 
 10.1093/mnras/stae845, 2024RAA....24e5015D}) 
are a class of galaxies characterized by a central surface brightness ($\mu_0$) 
fainter than 23.0 mag arcsec$^{-2}$ in B-band \citep{1997ARA&A..35..267I}.
The star formation rate (SFR) of LSBGs is lower than that of main-sequence galaxies \citep{2019ApJS..242...11L}. 
LSBGs are metal poor \citep{2023ApJ...948...96C} and lack of molecular gas \citep{2017AJ....154..116C}. 
The surface density of LSBGs are lower than that of high surface brightness galaxies \citep{1996ApJ...469L..89D}.

Spectroscopic data for LSBG are still very scarce.
Previous studies are mainly based on the fiber or long-slit spectroscopic data,
which cannot represent whole galaxies \citep{1994ApJ...426..135M, 2010A&A...520A..69Z, 2015MNRAS.454.3664B, 2017ApJ...837..152D, 2020ApJ...899L..12R}.
Integral field spectroscopy (IFS; \citealt{ALLINGTONSMITH2006244, 2008SPIE.7018E..2NV})
enables the simultaneous acquisition of spectral and spatial information across an observed object. 
Advances in IFS instruments, including the Palomar Cosmic Web Imager (PCWI), 
the Keck Cosmic Web Imager (KCWI), 
the Gemini Multi-Object Spectrograph (GMOS) 
and the Multi Unit Spectroscopic Explorer (MUSE),
have given rise to IFS sky surveys, exemplified by the Mapping Nearby Galaxies and Apache
Point Observatory survey (MaNGA; \citealt{2015ApJ...798....7B}).
IFS observations of LSBGs are highly valuable but remain limited \citep{2021ApJ...909...20S, 2024A&A...686A.247J}.

The molecular gas content of LSBGs has been discussed since 
\cite{1990AJ....100.1523S}. 
Low-J CO is detected in a handful of giant LSBGs and edge-on LSBGs, such as UGC 2082, UGC 7321, NGC 4244, UGC 6614, 
and Malin 2, with the molecular gas masses ranging between 10$^6$ to 10$^8$ M$_\odot$ 
\citep{2000ApJ...545L..99O, 2001ApJ...549L.191M, 2006ApJ...651..853D, 2010A&A...523A..63D}.
In Malin 1, the molecular gas surface density ($\rm \Sigma_{\rm H_2}$) 
is reported to be lower than 0.3 M$_\odot$ pc$^{-2}$ \citep{2022ApJ...940L..37G}.
\cite{2010A&A...523A..63D} estimated the $\rm \Sigma_{\rm H_2}$ for Malin 2 to be 1.0 
M$_\odot$ pc$^{-2}$ based on the detection of the CO(2-1) emission line.
Upper limits for molecular gas in LSBGs are typically around 10$^7$ M$_\odot$,
The molecular gas content in LSBGs is lower than those star-forming galaxies 
at the similar stellar mass \citep{2017AJ....154..116C}.
Several factors have been proposed to explain the scarcity of molecular gas in LSBGs.
The low metallicity environments of LSBGs may lead to inefficient cooling,
impacting the formation of giant molecular clouds. Additionally, the low interstellar medium densities in LSBGs
make molecular clouds more susceptible to photodissociation \citep{2017AJ....154..116C}.
\cite{2022ApJ...940L..37G} suggests that the warm interstellar medium (ISM) of LSBGs may cause the lack of low-J CO detections.
Moreover, gas phase metallicity information is crucial for determining the CO-to-H$_2$ conversion factor  
(X$_{\rm CO}$, \citealt{2012MNRAS.421.3127N}), introducing uncertainty into molecular gas mass. 

AGC 102004 is an edge-on low surface brightness galaxy (ELSBG) at a redshift of 0.0179 
 \citep{2023ApJ...948...96C}, and displays an extended structure along the major 
axis ($\sim$ 33 kpc at a surface brightness of 25 mag arcsec$^{-2}$, \citealt{2023ApJS..269....3M}).
AGC 102004 is detected from ultraviolet to near-infrared bands (FUV, NUV, u, g, r, i, z, W1, and W2),
but undetected in WISE W3 and W4 bands.
The image data of AGC 102004 are mainly from large sky area surveys including GALEX, SDSS, DESI, and WISE.
AGC 102004 has SDSS 3$''$ fiber spectrum and ALFALFA HI spectrum.  
The basic parameters measured based on those archival data are listed in Table \ref{table:agc}.

In the remaining sections of the paper, 
we detail the IFS observation of H$\alpha$ and [{\ion{N}{II}}] emission lines conducted using PCWI \citep{2014ApJ...786..106M},
 as well as CO(2-1) emission observation carried by JCMT in Section\,2.
We analyze the anomalous H$\alpha$ gas disk of AGC 102004 in Section 3.
The distribution of gas-phase metallicity is presented in Section 4.
We discuss the possible a minor/mini merger in AGC 102004 in Section 5.
In Section 6, we estimate the upper limit of molecular gas in AGC 102004.
We summarize our conclusions in Section 7.
We adopt a $\Lambda$CDM cosmology with $\rm \Omega_{m}$ = 0.3, 
$\rm \Omega_{\Lambda}$ = 0.7, $\rm H_0$ = 70\,km\,s$^{-1}$\,Mpc$^{-1}$.

\begin{table*}[h!tb]
  \caption{Basic Parameters of AGC 102004} 
  \begin{tabular}{c c c c c c c c c c c} 
  \hline\hline
  & 1  & 2 & 3 & 4 & 5 & 6 & 7& 8& 9 & 10\\ 
  \hline
  Source & Redshift & Distance  & M$_*$  & M$_{HI}$  & W50  & 12+log(O/H) &  Diameter & Inclination & b/a & Scale Length \\
  &   &  &  &  &  &  & & &  & \\
        &  &Mpc &10$^{8}$ M$_{\odot}$  & 10$^{9}$ M$_{\odot}$  &   km s$^{-1}$   &    & arcminute  & degree  &    & kpc  \\
  \hline  
  &   &  &  &  &  &  & & &  & \\ 
  AGC 102004 & 0.0179& 75.3&5.49  & 4.57  &  204   &     8.36       &     1.51       &      84      &  0.15  & 6.65 \\
             &       & $\pm$2.2 &   & $\pm$ 0.55 & $\pm$ 2 & $\pm$ 0.007  & & &  & \\ 
  \hline
  \end{tabular}
  \begin{tablenotes} 
       \footnotesize
       \item{Col 1: The redshift is from the Siena Galaxy Atlas 2020 (SGA-2020, \citealt{2023ApJS..269....3M});}
       \item{Col 2: The distance is from ALFALFA \cite{2018ApJ...861...49H};}
       \item{Col 3: The stellar mass is from the MPA-JHU catalog \citep{2003MNRAS.341...33K};}
       \item{Col 4: The HI gas mass is form ALFALFA \citep{2018ApJ...861...49H};} 
       \item{Col 5: The velocity width of the HI line profile measured at the 50\% level of
       each of the two peaks is from ALFALFA \citep{2018ApJ...861...49H};}
       \item{Col 6: The gas phase metallicity is derived by [{\ion{N}{II}}]$\lambda$6583/H$\alpha$ diagnostic from SDSS 3$''$ fiber spectrum \citep{2023ApJ...948...96C};}
       \item{Col 7: The diameter at the 25 mag arcsec$^{-2}$ surface brightness isophote (in optical) is from SGA-2020 \citep{2023ApJS..269....3M};}
       \item{Col 8: The inclination is from \cite{2011ApJS..196...11S};}
       \item{Col 9: The semi-minor to semi-major axis ratio is from SGA-2020 \citep{2023ApJS..269....3M};}
       \item{Col 10: The scale length is measured from SDSS g-band fitted by GALFIT edge-on disk profile \citep{2023ApJ...948...96C};}
 \end{tablenotes}
  \label{table:agc} 
\end{table*}

\section{observations and Data Reduction}
\subsection{The IFS Observations and Data Reduction}

The IFS observations of H$\alpha$ and [{\ion{N}{II}}] were carried out with PCWI
on the night of 2023 September 23 through the project CTAP2023-A0053 (PI: T. Cao).
We applied the Red grating (640-770nm), r$'$/Red filter, and the 45nm unmasked spectral bandpass.
The center wavelength for AGC 102004
is 6680 $\rm \AA$ corresponding the redshifted H$\alpha$.
PCWI has a spectral resolution (R=$\rm \lambda$/$\rm \Delta\lambda$) 
of 5000, and the field of view (FOV) is 60$''$ $\times$ 40$''$.

Targets were positioned on the diagonal in the FOV to cover a larger area of galaxies. 
The single exposure time was 20 minutes, and the total on-source integration time was 4.83 hours. 
An off-target blank sky background was observed every 1-2 hours, and a standard star was included. 
The average seeing on the night was $\sim$1.1$''$, and the instrument spatial resolution was 2.5$''$ $\times$ 1.1$''$.

We employed the standard CWI pipeline \citep{2014ApJ...786..106M} for data reduction 
and utilized CWITools \citep{2020arXiv201105444O} to correct the WCS, coadd the WCS-corrected data, and subtract residual background. 
The final spectral cube has a channel width equivalent to
10.05 km s$^{-1}$. The mean channel noise ($\sigma_{ch}$) is 4.5 $\times$ 10$^{-19}$ erg s$^{-1}$ cm$^{-2}$ $\rm \AA^{-1}$. 
The flux images of H$\alpha$ (an integration over the wavelength from 6677.52 to 6684.78 $\rm \AA$) and [{\ion{N}{II}}] 
(an integration over the wavelength from 6699.30 to 6704.58 $\rm \AA$) have noises of 6.0 $\times$ 10$^{-19}$ and 5.6 $\times$ 10$^{-19}$ erg s$^{-1}$ cm$^{-2}$, respectively.

\subsection{The CO(2-1) observations and Data Reduction}

The CO(2-1) observations of AGC 102004 were conducted using the $^\backprime\bar{\rm U}^\backprime\bar{\rm u}$ 
receiver mounted within the N$\bar{\rm a}$makanui instrument on JCMT through the project M22AP008 (PI: T. Cao). 
We employed the beam switch observation mode to obtain a stable baseline with a lower noise.
The half-power beam width of $^\backprime\bar{\rm U}^\backprime\bar{\rm u}$ is 20$''$ at 230.538 GHz.
The beam effeciency (B$\rm_{eff}$) is 0.66 for $^\backprime\bar{\rm U}^\backprime\bar{\rm u}$, 
and T$_{mb}$= T$^*_{\rm A}$/B$\rm_{eff}$.
Utilizing a 1 GHz bandwidth with 2048 channels, with a velocity resolution is 0.63 km s$^{-1}$ (one channel),
each observation took 40 minutes and we repeated 11 times for a total observation time of 7.33 hours.
Six observations (4.0 hours) were carried out under Band 4/5 weather conditions on May 28 and July 30, 2022. 
The remaining five observations (3.33 hours) were carried out under Band 2/3 weather conditions on December 26 and 27, 2023.

\begin{figure}
  \begin{center}
  \includegraphics[width=3.6in]{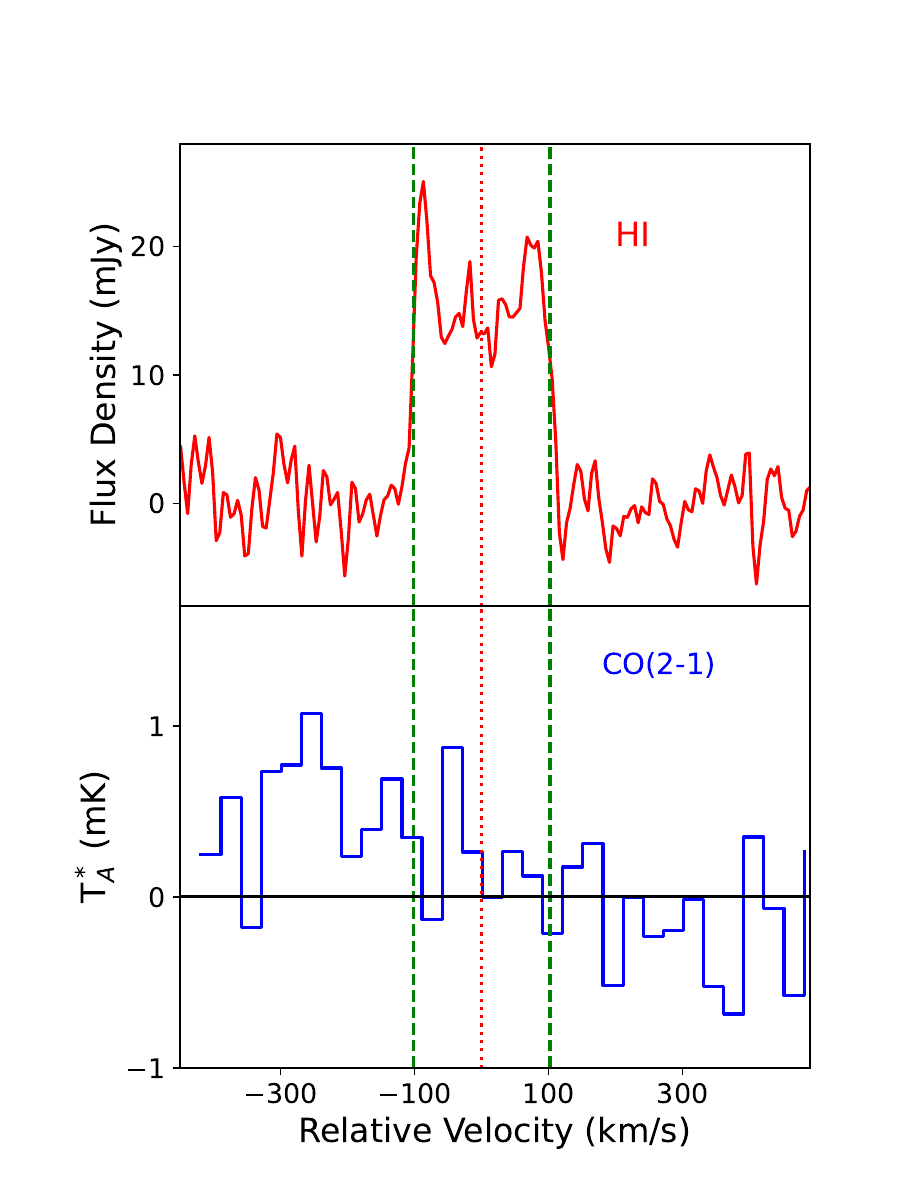}
  \end{center}
\caption{Upper Panel: The HI spectrum of AGC 102004 from ALFALFA \citep{2018ApJ...861...49H}.
Lower Panel: The combined spectrum of CO(2-1) of AGC 102004.
The spectrum is smoothed to 30 km s$^{-1}$ with 1$\sigma$ = 0.33 mK. 
The dotted line represents the relative velocity at zero.
The width of W50 (204 km/s$\pm$2 km s$^{-1}$, \citealt{2018ApJ...861...49H}) of HI line 
marked by two green dashed lines.}
           \label{A10CO}
\end{figure}

We processed the data using the Starlink software package ORAC-DR pipeline \citep{2015MNRAS.453...73J}, 
and converted the spectrum to GILDAS/CLASS format for further data processing. 
The $\sigma$, root mean square (RMS) noise, is 2.3 mK at native velocity resolution.
We smoothed the spectrum to 30 km s$^{-1}$ with $\sigma$ = 0.33 mK as shown in Figure\,\ref{A10CO}.
The CO(2-1) has non-detection in AGC 102004.

We summarize of our observations in Table \ref{table:obs}.

\begin{table*}[h!tb]
  \caption{Observations of AGC 102004} 
  \begin{tabular}{c c c c c c c} 
  \hline\hline
   Telescope & Instrument & observation time  & total on-source time & rms & emission line & beam size\\
  \hline 
   & & & & & &\\
    P200 &  CWI &  2023-09-23 &  4.83 hours &  4.5 $\times$ 10$^{-19}$ & [{\ion{N}{II}}]$\lambda$6583, H$\alpha$ & 2.5$''$$\times$1.1$''$ \\
    & & &  & erg s$^{-1}$ cm$^{-2}$ $\rm \AA^{-1}$& \\
    & & & & & &\\
    \hline  
    & & 2022-05-28 & & & &\\
    JCMT & $^\backprime\bar{\rm U}^\backprime\bar{\rm u}$ &  2022-07-30 & 7.33 hours & 2.3 mK & CO(2-1) & 20$''$ \\
    & & 2023-12-26& & at 0.63 km/s  & &\\
    & & 2023-12-27& & & &\\
  \hline
  & & & & & &\\
 & & & Results & & & \\
 & & & & & &\\
 \hline\hline 
  I$\rm _{CO_{(2-1)}}$ & L$\rm _{CO}$  & M$\rm _{H_2}$$^{\textbf{A}}$ & M$\rm _{H_2}$/M$^{\rm corr}\rm _{HI}$ &  M$\rm _{H_2}$/M$\rm _{*}$ & SFR$\rm _{H\alpha}$ & M$\rm _{H_2}$/SFR$\rm _{H\alpha}$ \\
  & & & & & &\\
  K km s$^{-1}$ & 10$^6$ K km s$^{-1}$ pc$^{-2}$ & 10$^7$ M$\odot$ &  &   &  M$\odot$ yr$^{-1}$ & Gyr \\
  \hline
  & & & & & &\\
  $<$ 0.116  & $<$ 4.4 & $<$ 2.1 & $<$0.0037  & $<$ 0.038  & 0.065$\pm$0.0013 & $<$ 0.3 \\
  & & & & & &\\
  \hline
  \end{tabular}
  \begin{tablenotes} 
    \footnotesize
    \item{
    A: The molecular mass is derived from the best-fit value of X$_{\rm CO}$ of 3.02$\times$10$^{20}$ cm$^{-2}$ (K km s$^{-1}$)$^{-1}$ at 0.4 Z$_\odot$ \citep{2012MNRAS.421.3127N}.}
  \end{tablenotes}
  \label{table:obs} 
\end{table*}

\begin{figure}
  \begin{center}
  \includegraphics[width=3.3in]{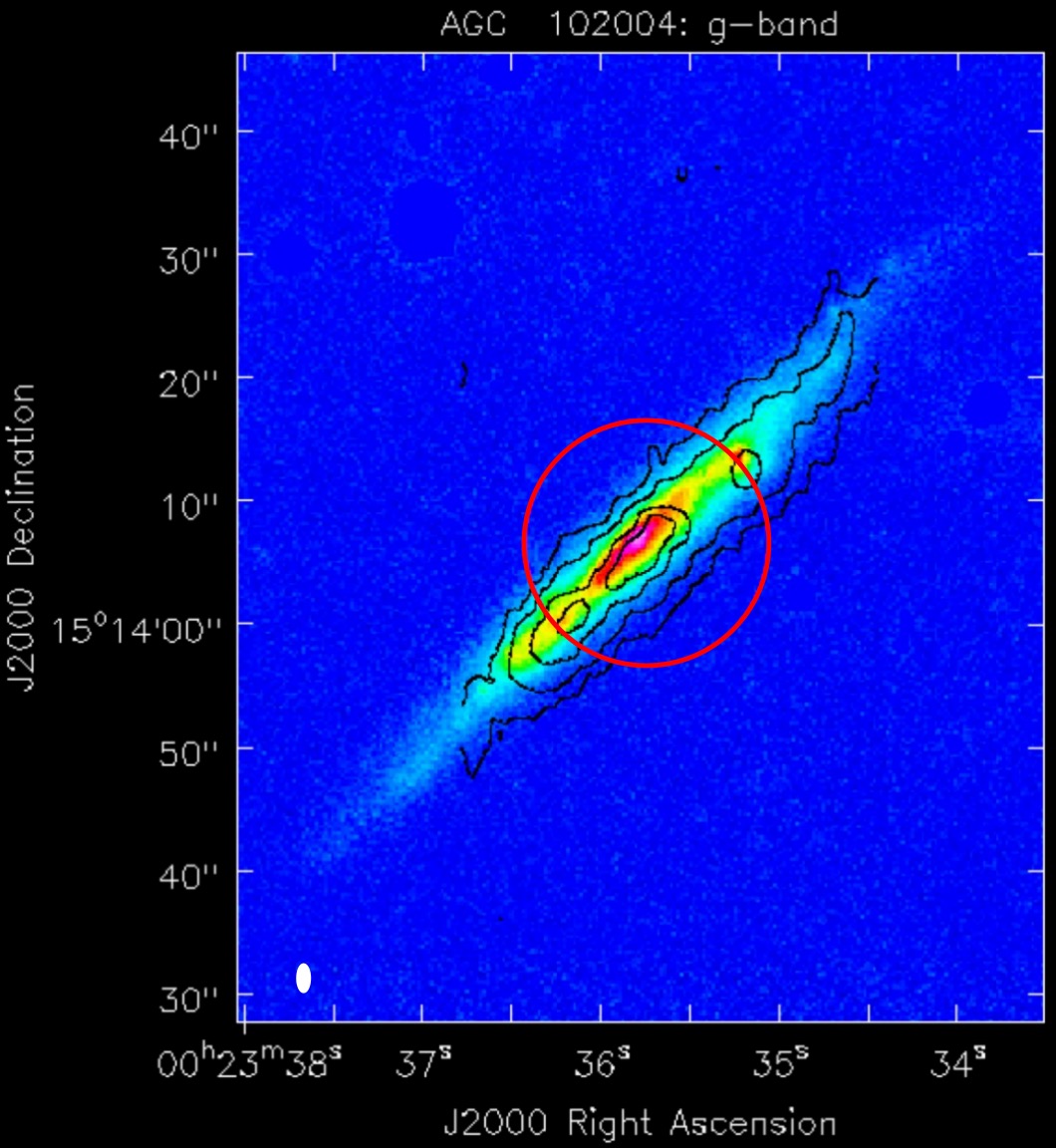}
  \end{center}
\caption{
The beam of JCMT 15-meter telescope for CO(2-1) (red circle) overlaid on the DESI g-band image of AGC 102004.
Integrated H$\alpha$ emission contours (black line) at [3, 10, 22, 36] $\times$ $\sigma$ 
(where $\sigma$ = 6.0 $\times$ 10$^{-19}$ erg s$^{-1}$ cm$^{-2}$) overlaid on DESI g-band image of AGC 102004. 
The images are produced by CASA software \citep{2007ASPC..376..127M}. 
The oval in the lower left corner represents the spatial resolution of H$\alpha$ (2.5$''$ $\times$ 1.1$''$).}
           \label{A10war}
\end{figure}

\section{The results from ISF observations of H$\alpha$}

We present integrated H$\alpha$ line emission 
contours of AGC 102004 overlaid on DESI g-band image in Figure\,\ref{A10war}. 
The overall morphology of the H$\alpha$ gas disk is consistent with its g-band morphology. 
The incomplete H$\alpha$ coverage of our observation is evident in the g-band.
We apply GALFIT \citep{2002AJ....124..266P} S$\acute{e}$rsic profile to define the major axis and center 
position of the H$\alpha$ image, with a Position Angle (PA) of 49$^{\circ}$. 
We correct our observed velocity frame to the Heliocentric reference frame by 4.8 km s$^{-1}$.
The Gaussian fit center velocity is 5371.2 km s$^{-1}$ of H$\alpha$ in the Heliocentric velocity system.
We note we use the Heliocentric reference frame and optical velocity definition in the following of the paper.

\subsection{ H$\alpha$ velocity distribution in the NW disk}
\begin{figure*}
  \begin{center}
  \includegraphics[width=7in]{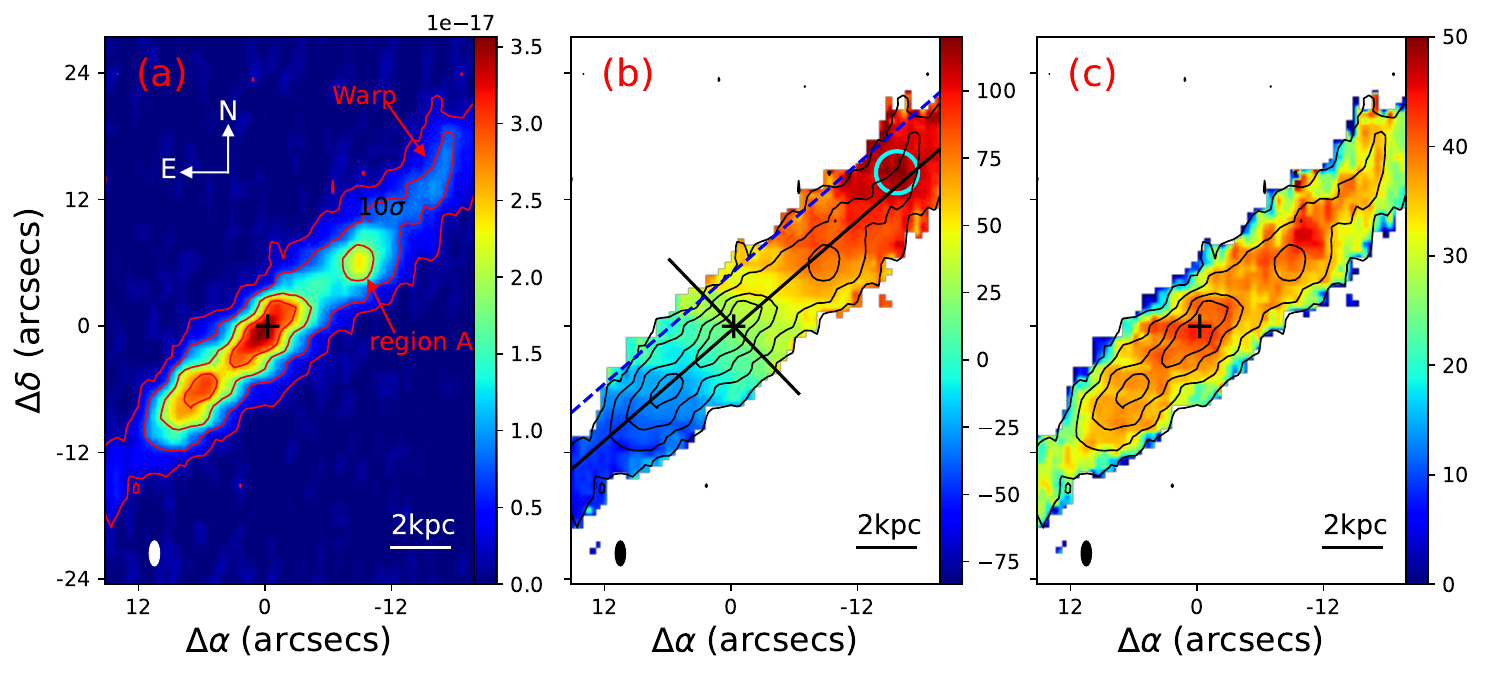}
  \end{center}
\caption{Integrated H$\alpha$ contours at [3, 10, 22, 36] $\times$ $\sigma$ 
(where $\sigma$ = 6.0 $\times$ 10$^{-19}$ erg s$^{-1}$ cm$^{-2}$) overlaid on the images of
same integrated H$\alpha$ line emission image in (a), the H$\alpha$ emission line velocity field (moment 1) in (b), 
and the velocity dispersion (moment 2) in (c). 
The images in (b) and (c) are generated by those spaxels above 3$\sigma_{ch}$, where $\sigma_{ch}$ is the mean channel noise 
 ($\sigma_{ch}$= 4.5 $\times$ 10$^{-19}$ erg s$^{-1}$ cm$^{-2}$ $\rm \AA^{-1}$) of AGC 102004. 
We mark the major axis and minor axis by black lines in panel (b). 
The blue dashed line parallels to major axis and at 4.0$''$ offset from the perpendicular disk center.
The oval in the lower left corner represents the spatial resolution 2.5$''$ $\times$ 1.1$''$.}
           \label{A10vel}
\end{figure*}

The integrated H$\alpha$ line emission image of AGC 102004 in Figure\,\ref{A10vel} (a)
reveals an intensity peak in the central region,
with two sub-peaks distributed in the SE and NW disks, respectively. 
The sub-peak in the NW is weaker than that in the SE.
Notably, our observation does not cover the full steller disk of AGC 102004 in Figure\,\ref{A10war},
and the H$\alpha$ disk along the major axis extending up to 7 kpc and 9 kpc for the SE and NW disks, respectively.
The warp structure in the NW disk is evident in Figure\,\ref{A10vel} (a) 
as indicated by the arrow.
Figure\,\ref{A10vel} (b) illustrates the H$\alpha$ velocity field.
An anomalous velocity component is observed around the warp structure in the NW disk, 
as indicated by the cyan circle in Figure\,\ref{A10vel} (b).
Additionally, Figure\,\ref{A10vel} (c) displays the velocity dispersion ($\sigma_{v}$)
within 3$\sigma$ H$\alpha$ contour. 
In the NW disk, near region A, the $\sigma_{v}$ is slightly larger than that in the center region. 
The $\sigma_{v}$ shows a relatively high value along the region A to the warp structure.

\begin{figure}
  \begin{center}
  \includegraphics[width=3.0in]{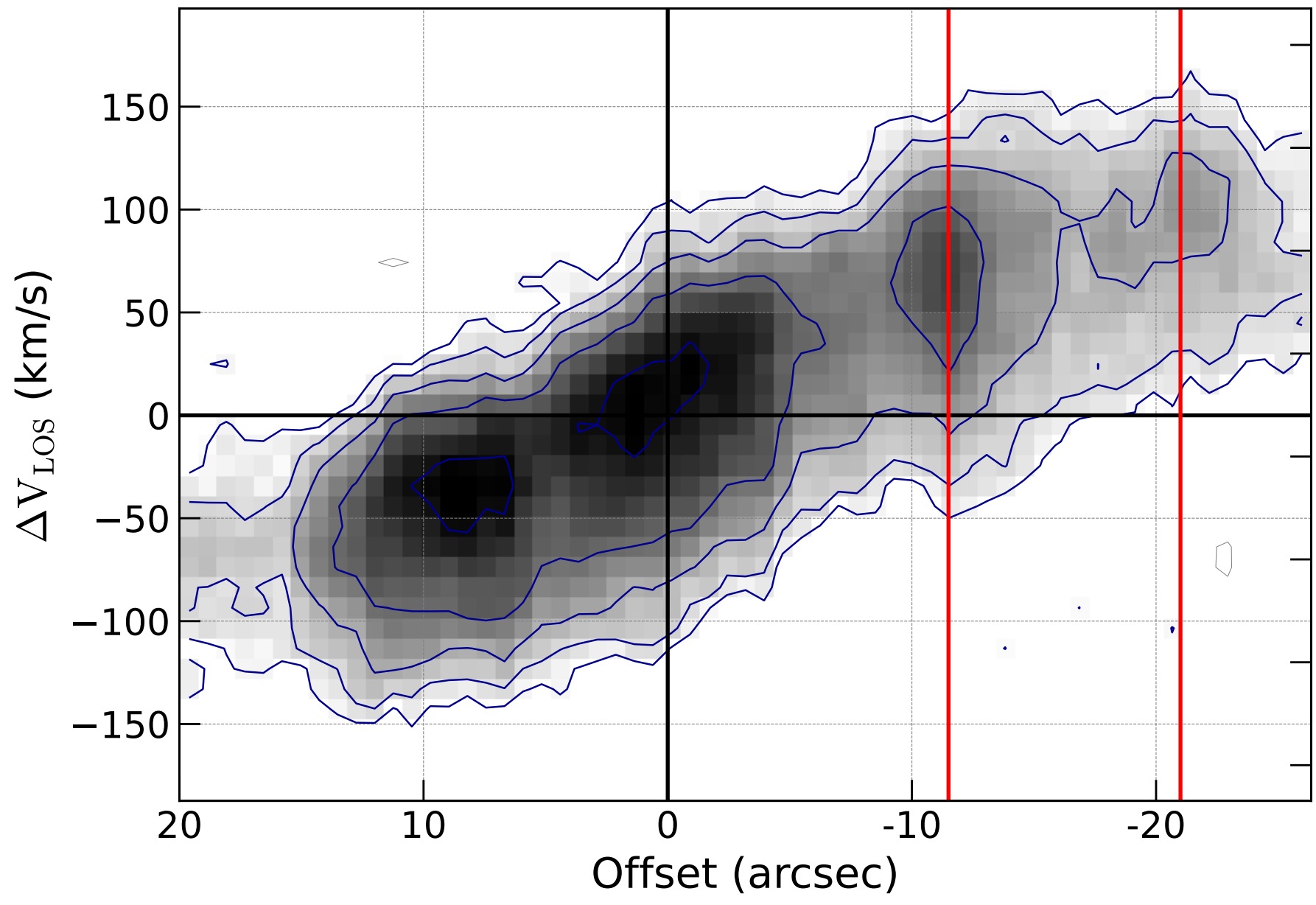}
  \end{center}
\caption{The H$\alpha$ PVD along the major axis of AGC 102004. 
H$\alpha$ is shown in grey with blue contours.
The red lines mark the region A and warp structure position in Figure\,\ref{A10vel} (a) along the major axis.}
           \label{pv_1}
\end{figure}

We utilize $^{3D}$BAROLO \citep{2015MNRAS.451.3021D} to generate the Position-velocity Diagram (PVD).
Figure\,\ref{pv_1} displays PVD along the major axis. 
Two red lines are positioned at offsets of -11.5$''$ (equivalent to 4.20 kpc) 
and -21.0$''$ (equivalent to 7.66 kpc) from the center,
indicating the locations of region A and the warp structure along the major axis as depicted in Figure\,\ref{A10vel} (a).
The projected physical size between the two red lines is 3.46 kpc. 
The NW disk presents different structure from the SE disk in PVD.

\subsection{Spectra and channel map of H$\alpha$}
\begin{figure}
  \begin{center}
  \includegraphics[width=3in]{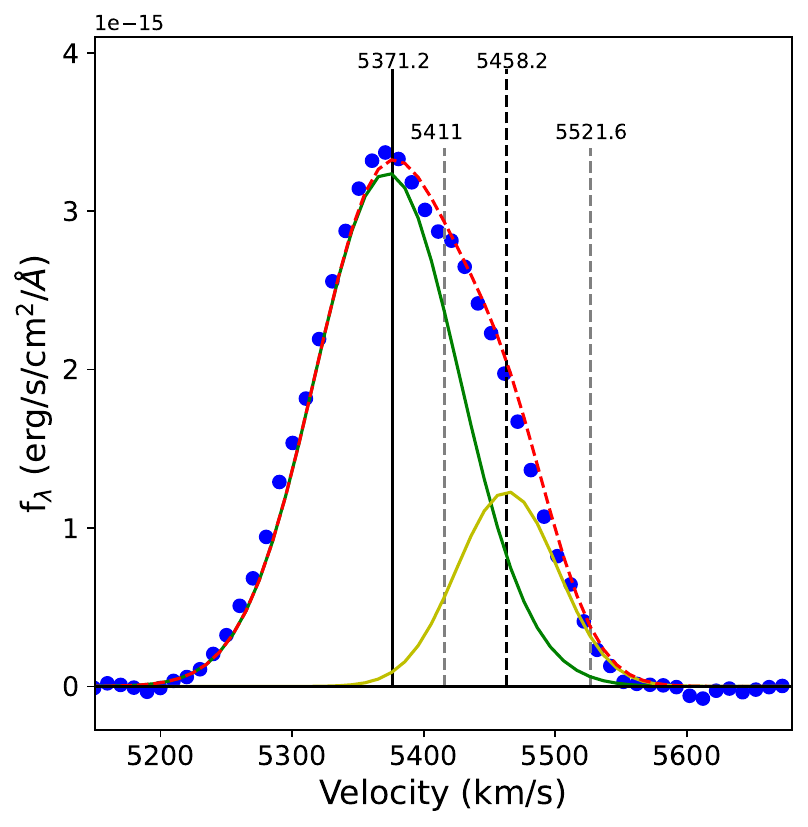}
  \end{center}
\caption{H$\alpha$ spectrum of AGC 102004 among 
3$\sigma$ contour region in Figure\,\ref{A10vel} (a). 
The blue dots are the observed data and the red dashed line is the Gaussian fit contain two Gaussian components 
(as green and yellow lines shown). The dashed black line is the center velocity of the yellow component. Two grey
dashed lines mark velocities at 5411 and 5521.6 km s$^{-1}$. 
The black line at 5371.2 km s$^{-1}$ is the center velocity of Gaussian fit (the red dashed line).}
          \label{spectra}
\end{figure}

\begin{figure*}
  \begin{center}
  \includegraphics[width=6in]{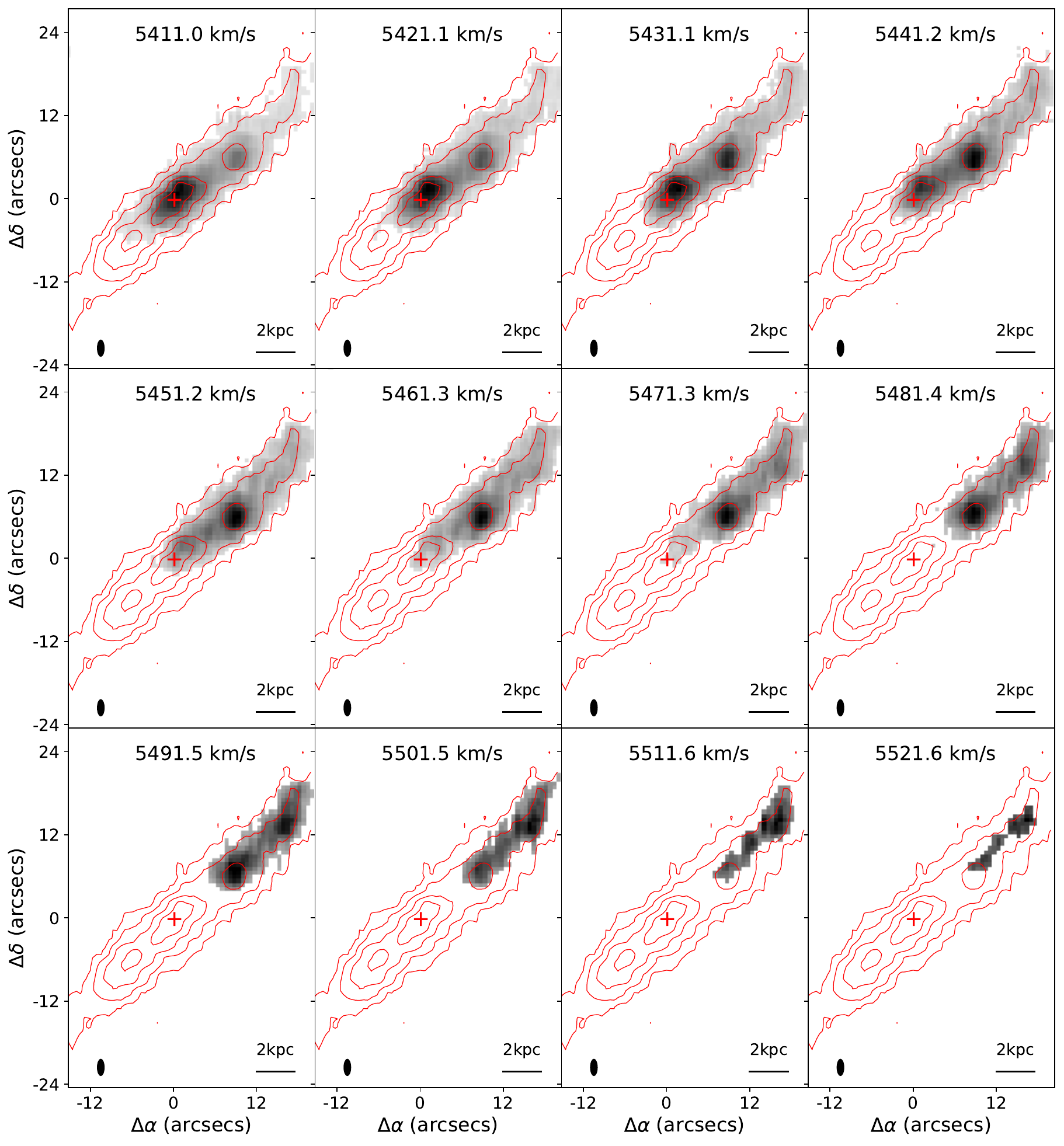}
  \end{center}
\caption{H$\alpha$ Channel maps (from 5411 to 5521.6 km s$^{-1}$) of AGC 102004. The red contours are the total flux-integrated H$\alpha$ emission 
(e.g., from Figure\,\ref{A10vel} (a)). The images are only using spaxels above 3$\sigma_{ch}$, where $\sigma_{ch}$ is the mean channel noise 
($\sigma_{ch}$= 4.5 $\times$ 10$^{-19}$ erg s$^{-1}$ cm$^{-2}$ $\rm \AA^{-1}$) of AGC 102004.
The oval in the lower left corner represents the spatial resolution 2.5$''$ $\times$ 1.1$''$.}
           \label{channel}
\end{figure*}

Figure\,\ref{spectra} displays the H$\alpha$ spectrum of AGC 102004 within
3$\sigma$ contour region outlined in Figure\,\ref{A10vel} (a), 
and the spectrum is well-fitted by two Gaussian components.
The prominent Gaussian component (green line) is centered at 5367.4$\pm$54.76 km s$^{-1}$, 
while the weak Gaussian component (yellow line) is centered at 5458.2$\pm$38.28 km s$^{-1}$. 
In Figure\,\ref{channel}, channel maps spanning from 5411 to 5521.6 km s$^{-1}$ are presented, 
as indicated by the grey dashed lines in Figure\,\ref{spectra}.
The channels at 5491.5, 5501.5, 5511.6 and 5521.6 km s$^{-1}$ 
(the bottom panels in Figure\,\ref{channel}) reveal
distinct morphologies along region A and the warp structure in NW disk.
These channels are mainly contributed by the weak Gaussian component (see Figure\,\ref{spectra})
and along the warp structure.

\begin{figure}
  \begin{center}
  \includegraphics[width=3.0in]{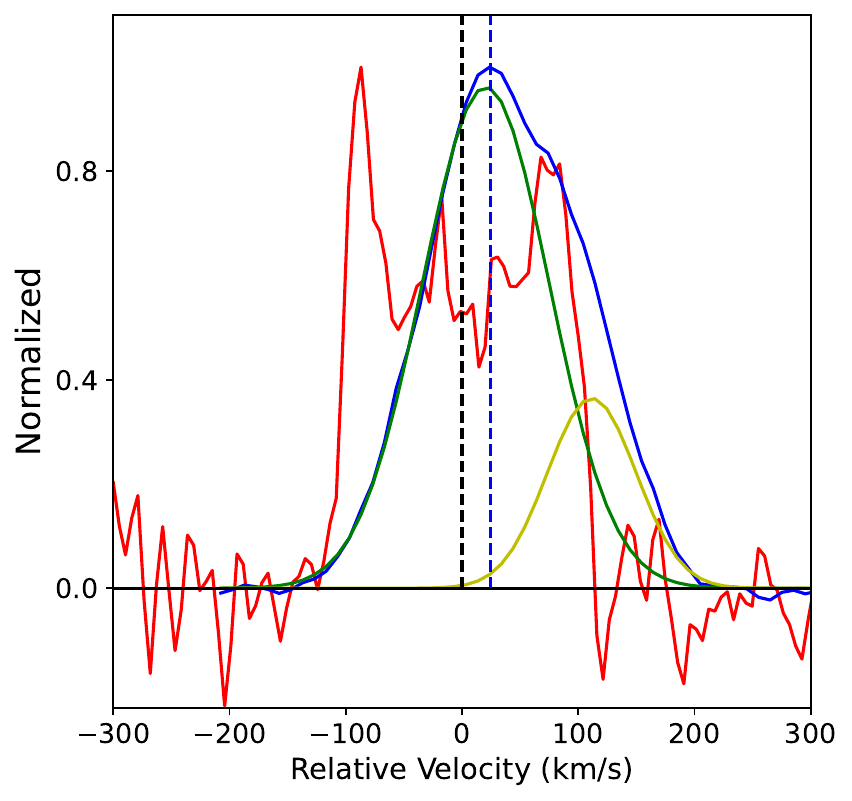}
  \end{center}
\caption{The red line is the HI spectrum from ALFALFA, and the blue line is the 
H$\alpha$ spectrum of AGC 102004 among 3$\sigma$ contour region in Figure\,\ref{A10vel} (a).
The green and yellow solid lines are two Gaussian components in Figure\,\ref{spectra}.
The black dashed line is the systemic velocity of HI emission.
The H$\alpha$ emission center of Gaussian fit (blue dashed line) is offset by +24.2 km s$^{-1}$ from the 
systemic center velocity obtained from the HI emission.}
           \label{HI}
\end{figure}

We compare our H$\alpha$ spectra to HI spectra from ALFALFA in Figure\,\ref{HI}.
The HI spectrum displays a horn structure and the NW peak appears to be broken.
However, we cannot ascertain if this is related to noise.
We did not derive all of the flux from the SE disk because of the limit of our observation field of view, 
as mentioned in Section 3. 
But it may not impact much of the H$\alpha$ spectrum since 
the H$\alpha$ signal is very weak in the outskirt.
There is an evident velocity excess of H$\alpha$ compared to HI in the NW wing,
corresponding to the anomalous velocity of the warp structure in the NW disk, 
and the center of H$\alpha$ emission of Gaussian fit is offset by +24.2 km s$^{-1}$ from the 
systemic velocity obtained from the HI emission. We note the channel spacing of 
HI spectrum from ALFALFA is 5.1 km s$^{-1}$ \citep{2018ApJ...861...49H}.

\section{Metallicity distribution in AGC 102004}

\begin{figure*}
  \begin{center}
  \includegraphics[width=7.0in]{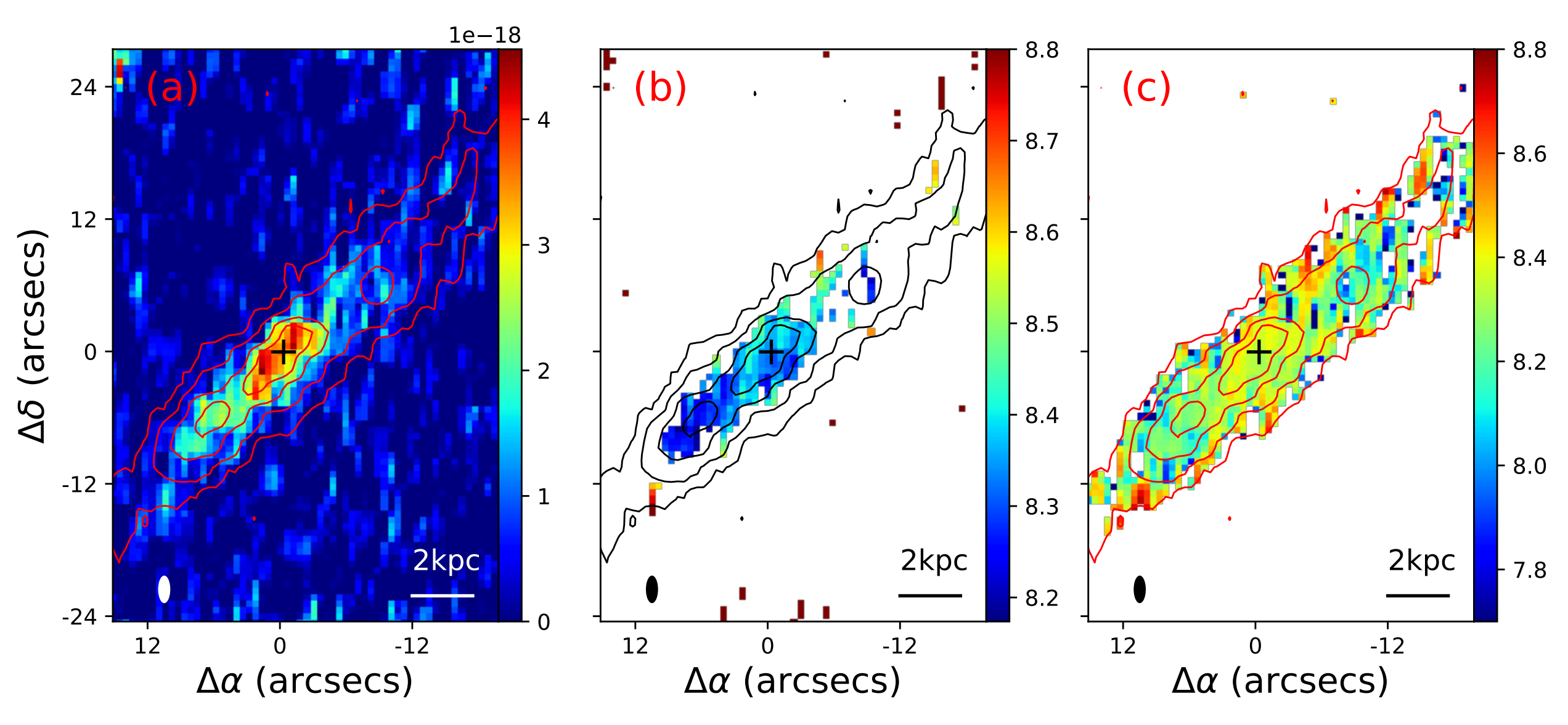}
  \end{center}
\caption{Integrated H$\alpha$ contours at [3, 10, 22, 36] $\times$ $\sigma$ 
(where $\sigma$ = 6.0 $\times$ 10$^{-19}$ erg s$^{-1}$ cm$^{-2}$) overlaid on the
integrated [{\ion{N}{II}}] line emission image in Panel (a) and the gas-phase abundance 12+log(O/H) in Panels (b) and (c).
12+log(O/H) is derived Panel (b) is only using spaxels above 3$\sigma_{[NII]}$
(1 $\sigma_{[NII]}$ = 5.6 $\times$ 10$^{-19}$ erg s$^{-1}$ cm$^{-2}$). 
Panel (c) included those low signal spaxels. 
The oval in lower left corner represents the spatial resolution 2.5$''$ $\times$ 1.1$''$.
}
           \label{A10all}
\end{figure*}

Panel (a) of Figure\,\ref{A10all} presents the [{\ion{N}{II}}] line emission, 
and we derived the metallicity 12+log(O/H) in panel (b) using 
[{\ion{N}{II}}]$\lambda$6583/H$\alpha$ ([N2] index \citealt{2004MNRAS.348L..59P}). 
The metallicity estimation considered only those spaxels above 3$\sigma$ of 
[{\ion{N}{II}}] in Figure\,\ref{A10all} (b). 
Metallicities range from 8.15 to 8.74, and 
the mean value of 12+log(O/H) is 8.36 $\pm$ 0.10.

We further assessed the metallicity of regions with low signal [{\ion{N}{II}}] region within 
3$\sigma$ H$\alpha$ contour region.
In Figure\,\ref{A10all} (c), metallicities range from 7.72 to 8.77, and 
the mean 12+log(O/H) is 8.30 $\pm$ 0.19.
We note that the 3$\sigma$ [{\ion{N}{II}}] detection is predominantly in the center and SE disk, 
and 12+log(O/H) values of the NW disk represent the upper limits of metallicity.
Furthermore, the metallicity of the outer disk is lower than the detection limit which is 7.72.
This suggests that the metallicity gradient exists in AGC 102004.

\cite{2015MNRAS.454.3664B} identified the metallicity gradients as a common feature of LSBGs by 
examining the spectra of 141 {\ion{H}{II}} regions in ten LSBGs.
\cite{2017MNRAS.469..151B} demonstrated a notably flat gradient (almost constant) for low-mass galaxies 
(M$_*$ $\sim$ 10$^9$ M$_\odot$) using MaNGA IFS results.
Recent studies suggest that low metallicity with a substantial gradient may exist in merger systems or gas accretion galaxies, 
exemplified by cases like the NGC\,4809/4810 merger system\citep{2023A&A...677A.179G} and MaNGA 8313-1901 \citep{2022ApJ...938...96J}. 
Metallicity gradients in LSBGs may signify the occurrence of mergers or metal-poor gas accretion events.

\section{A possible merger event in AGC 102004}

There are no massive galaxies around AGC 102004. 
The two nearest neighbors (at a similar redshift) are located at approximately 10 
arcminutes away from AGC 102004 ($\sim$219 kpc in projected distance). 
One neighbor is in the northwest direction and the other is in the south direction.
No obvious tidal tail or stellar shell structures are observed in two nearest neighbors 
of AGC 102004 by DESI images.

The merger event could result in the observed warp structure in the NW disk 
and it also could generate turbulence to trigger star-forming activity in region A.
This turbulence could be reflected by the anomalous velocity in the NW disk
since the merger can bring additional momentum and disturbence and destroy
the original stable state. 
The metallicity gradient in AGC 102004 can support the merger event.
It is more likely a minor/mini merger in AGC 102004, 
because major mergers should leave some obvious traces.

Mergers can exert a substantial influence on galaxies 
as demonstrated by studies such as \cite{2022ApJ...934..114Y} and \cite{2023ApJ...949...83H}.
Most merger topics are on those high mass and high SFR counterparts. 
We highlight that the observational evidence of merger events in dwarf-mass galaxies is still scarce 
(e.g. Dwarf Galaxy VCC 848, \citealt{2020ApJ...891L..23Z, 2020ApJ...900..152Z}).
How the merger influences those low-mass galaxies is not clear. 
\cite{10.1088/1674-4527/ad5399} investigated the formation of superthin galaxies in the TNG100 simulation
and found that superthin galaxies undergo high frequency of prograde mergers since z $\sim$ 1.
We note that superthin galaxies are LSBGs seen in edge-on \citep{2013MNRAS.431..582B, 2022MNRAS.514.5126N}. 
Our observation results could potentially serve as observational evidence that mergers exist in LSBGs.

Additionally, metal-poor gas inflow, and misalignment between stellar disks 
and dark matter halos \citep{2022ApJ...935...48Z} can result the warp structure in galaxy.
More data are needed to further investigation for those scenarios in AGC 102004.

\section{The molecular gas content in AGC 102004}

\subsection{The upper limit of molecular gas mass}

The upper limit of integrated CO(2-1) flux density is drived by 3$\sigma$$\sqrt{\delta v \times \Delta V}$/B$\rm_{eff}$ \citep{2020MNRAS.499L..26W},
where the $\sigma$ is the rms noise at native velocity resolution and $\delta$v is the native velocity resolution.
The line width $\Delta$V is adopted W50 width of HI emission.

We adopt the R$_{21}$ = CO(2-1)/CO(1-0)=0.7 \citep{2009AJ....137.4670L, 2012AJ....143..138S, 2017AJ....154..116C}. 
The CO-to-H$_2$ equation \citep{1997ApJ...478..144S,2020MNRAS.499L..26W} is expressed as follows:

\begin{equation}
  \rm M_{\rm H_2} = \alpha_{CO} \times L_{\rm CO} 
\end{equation}

\begin{equation}
  \rm L_{\rm CO} = 23.5 \times \frac{\pi}{4ln2} \times \rm\theta_{mb}^2 \times I_{\rm CO} \times D^2 \times (1+z)^{-3}
  \end{equation}

Here, M$_{H_2}$ is in unit M$_\odot$ and L$_{\rm CO}$ is the CO luminosity in K km s$^{-1}$ pc$^{-2}$.
$\alpha_{CO}$ is given in M$_\odot$ (K km s$^{-1}$ pc$^{-2}$)$^{-1}$ \citep{2013ARA&A..51..207B}, 
which equals X$_{\rm CO}$/(6.3$\times$10$^{19}$ pc$^2$ cm$^{-2}$ M$_\odot^{-1}$).
$\rm\theta_{\rm mb}$ = 20$''$ is the observation beam size of JCMT $^\backprime\bar{\rm U}^\backprime\bar{\rm u}$ 
receiver.
The I$_{\rm CO}$ is integrated CO(1-0) flux density in K km s$^{-1}$.
 D is the distance in Mpc, and z is the redshift.

With a mean metallicity 12+log(O/H)$\sim$8.30
 ($\sim$ 0.4 Z$_\odot$; Z$_\odot$ = 8.76, \citealt{2011SoPh..268..255C}) of AGC 102004, 
we obtain a corresponding X$_{\rm CO}$ value 3.02$\times$10$^{20}$ cm$^{-2}$ (K km s$^{-1}$)$^{-1}$ 
at the best fit of the relationship between the 
X$_{\rm CO}$ to metallicity \citep{2012MNRAS.421.3127N}. 
The upper limits of M$_{\rm H_2}$ in AGC 102004 is 2.1 $\times$ 10$^7$ M$\odot$.

The X$_{\rm CO}$ factor ranges from 0.5 to 15.5 $\times$10$^{20}$ cm$^{-2}$ (K km s$^{-1}$)$^{-1}$
at 0.4 Z$_\odot$ in \cite{2012MNRAS.421.3127N} encompassing all model galaxies. 
If we use the largest X$_{\rm CO}$ = 15.5 $\times$10$^{20}$ cm$^{-2}$ (K km s$^{-1}$)$^{-1}$ at 0.4 Z$_\odot$, 
and M$_{\rm H_2}$ is about five times of above estimation.
Variations in the chosen X$_{\rm CO}$ value do not significantly change our results. 
We apply the CO luminosity in Figure\,\ref{A10CO1} to compare AGC 102004 with other galaxies avoiding the uncertainty
from X$_{\rm CO}$ factor.

\subsection{M$\rm _{H_2}$/M$\rm _{HI}$, M$\rm _{H_2}$/M$\rm _{*}$ and molecular gas depletion time}

We corrected the self-absorption of HI affecting the densest regions of galaxy disks using 
M$^{\rm corr}_{\rm HI}$ = (a/b)$^{0.12}$$\times$M$_{\rm HI}$ \citep{1994AJ....107.2036G, 2005ApJS..160..149S},
 where a and b are the optical major and minor axes.
The M$^{\rm corr}_{\rm HI}$ is 5.71 $\times$10$^9$ M$_\odot$ of AGC 102004.

The gas mass ratio of M$_{\rm H_2}$/M$^{\rm corr}_{\rm HI}$ is lower than 0.0037 in AGC 102004.
The gas mass ratio of M$_{\rm H_2}$/M$_{\rm HI}$ is 0.2 in typically brighter Sd-Sm spirals 
\citep{1989ApJ...347L..55Y} and in normal galaxies are from $\sim$0.03 to $\sim$3 
with a median value of $\sim$ 0.3 \citep{2015ApJ...799...92J}. 
We note \cite{2015ApJ...799...92J} adopted X$_{\rm CO}$ value 2.0$\times$10$^{20}$ cm$^{-2}$ (K km s$^{-1}$)$^{-1}$. 
The gas mass ratio of AGC 102004 is about two magnitude lower than the average level of normal galaxies.

The APEX low-redshift legacy survey for molecular gas
(ALLSMOG, \citealt{2014MNRAS.445.2599B, 2017A&A...604A..53C}) investigated CO in 
local low M$_*$ (10$^{8.5}$ $<$ M$_*$ $<$ 10$^{10}$ M$_\odot$) star-forming galaxies, 
and the mean M$_{\rm H_2}$/M$_*$ is 0.09-0.13 \citep{2014MNRAS.445.2599B}.
We note that ALLSMOG reached an uniform rms $\sim$ 0.8 mK with 50 km s$^{-1}$ for 
the APEX 230 GHz observations \citep{2017A&A...604A..53C}. 
The molecular gas fraction M$_{\rm H_2}$/M$_*$ of AGC 102004 is lower than 0.038.

We estimate the SFR by H$\alpha$ luminosity from our IFS observation 
following the method by \cite{1998ARA&A..36..189K} without dust correction. 
The derived SFR for AGC 102004 is 0.065 $\pm$ 0.0013 M$_\odot$ yr$^{-1}$, 
and the specific star formation rate (sSFR) is 1.3$\times$ 10$^{-10}$ yr$^{-1}$. 
Molecular gas depletion time $\rm \tau^{\rm H_2}_{\rm dep}$ = M$_{\rm H_2}$/SFR$_{\rm H\alpha}$ 
is shorter than 0.3 Gyr for AGC 102004,
which is shorter than that of the nearby disk galaxies 1-3 Gyr
\citep{2008AJ....136.2782L, 2013AJ....146...19L, 2017ApJ...845..133S}.

The inefficiency of star formation in AGC 102004 is likely caused by the low efficiency in converting molecules from atomic gas,
which is consistent with the discussions in \cite{2017AJ....154..116C} and \cite{2020MNRAS.499L..26W}.
All the above results are listed in Table\,\ref{table:obs}.

\subsection{Comparison with other galaxies}

\begin{figure*}
  \begin{center}
  \includegraphics[width=5.5in]{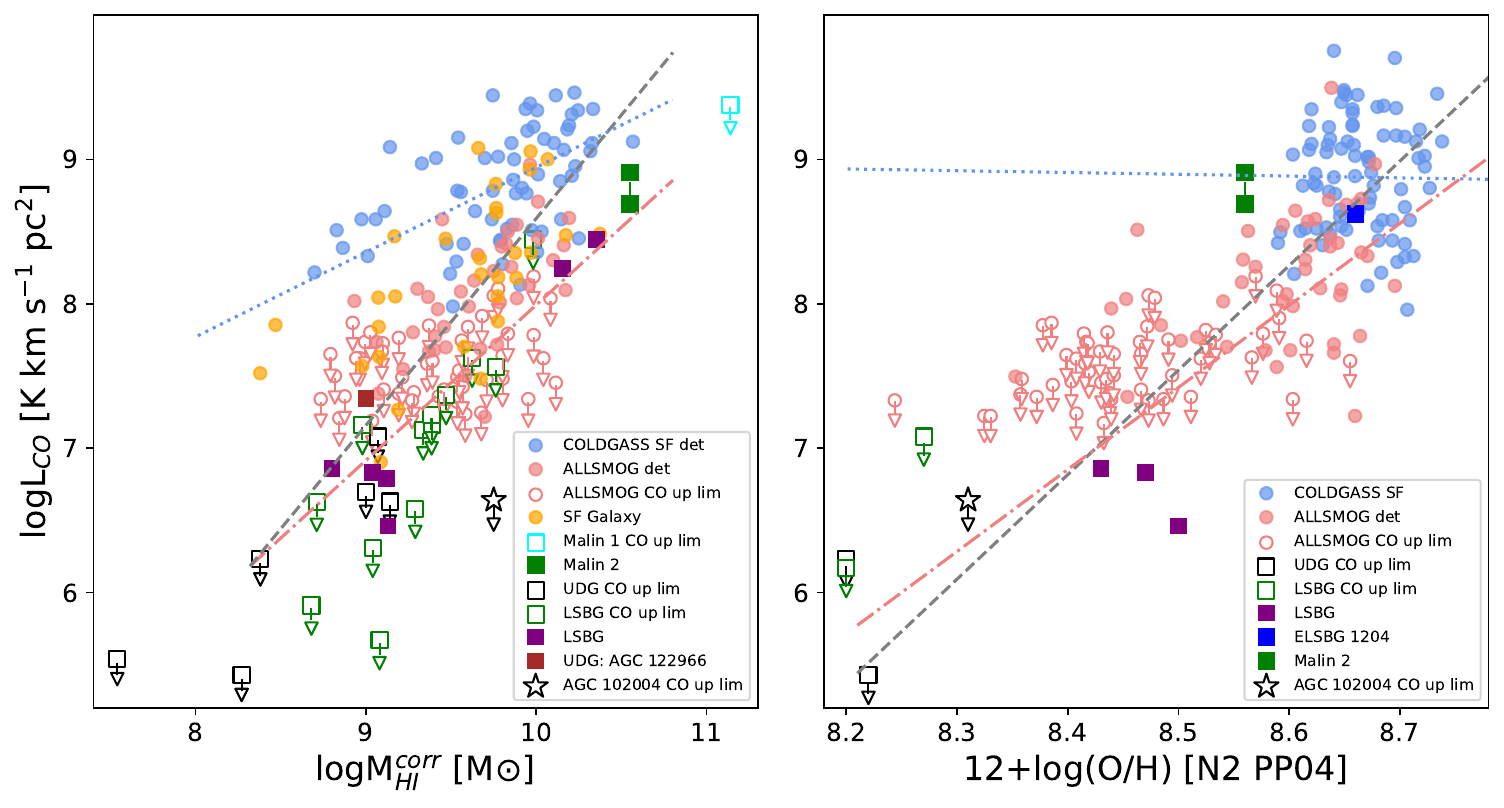}
  \end{center}
\caption{Left Panel: CO luminosity as a function of HI 
gas mass. The upper limits of CO luminosity of LSBGs (green unfilled squares) are from 
\cite{2001ApJ...549L.191M} and \cite{2017AJ....154..116C}. Cyan square respresents Malin 1 from \cite{2022ApJ...940L..37G}
and the green squares respresents Malin 2 from \cite{2010A&A...523A..63D}.
Six LSBGs with CO detection are from \cite{2001ApJ...549L.191M} and \cite{2003ApJ...588..230O}.
Ultra-diffuse galaxies (UDGs) are from \cite{2020MNRAS.499L..26W}.
The gas-rich star-forming (SF) galaxies are from \cite{2015ApJ...799...92J}.
Right Panel: CO luminosity as a function of 12+log(O/H).
The gas phase metallicity is estimated by the [N2] index.
The CLOD GASS and ALLSMOG data points, and 
the fit lines are from \cite{2017A&A...604A..53C}. 
The black star mark represents the upper limit of AGC 102004.
The LSBGs and UDGs are from \cite{2001ApJ...549L.191M} and \cite{2020MNRAS.499L..26W}.}
           \label{A10CO1}
\end{figure*}

We show the relationship between the CO luminosity and M$^{\rm corr}_{\rm HI}$ in the left panel of Figure\,\ref{A10CO1}.
The CO luminosity of AGC 102004 is significantly lower than that of the average value
of star-forming galaxies at similar M$^{\rm corr}_{\rm HI}$ mass.
Additionally, both LSBGs and UDGs (detection or upper limits) show lower CO luminosities 
than those star-forming galaxies.

In the right panel of Figure\,\ref{A10CO1},
we illustrate the relationship between CO luminosity and the gas phase metallicity estimated by the [N2] index.
The detection rate for ALLSMOG galaxies is 75\% at 12+log(O/H) $>$ 8.5.
\cite{2017A&A...604A..53C} discussed the dependence of CO emission on O/H, 
attributing it to a high CO photodissociation rate in low-metallicity environments 
\citep{2010ApJ...716.1191W, 2011MNRAS.412..337G}. 
Just few LSBGs/UDGs from \cite{2001ApJ...549L.191M} and \cite{2020MNRAS.499L..26W} have metallicity information in Figure\,\ref{A10CO1}.
If all galaxies are considered in the right panel, the CO luminosity seems to increase with metallicity increasing.
It may suggest that the low metallicity environments in LSBGs may contribute to the observed low detection rate of low-J CO lines.

There needs more CO detection data to confirm the relationship for LSBGs/UDGs in Figure\,\ref{A10CO1}.
  
\section{CONCLUSIONS}

In this study, we present new IFS observtions of ionized gas (H$\alpha$ and [{\ion{N}{II}}])  
and CO(2-1) emission observations of an edge-on low surface brightness galaxy AGC 102004. 
Our primary findings are summarized as follows:

(1) The H$\alpha$ gas disk is consistent with its g-band morphology.
The velocity field of H$\alpha$ of the galaxy shows an anomalous distribution in the NW disk. 
The H$\alpha$ spectrum is well-fitted by two Gaussian components, 
and the weak Gaussian component is dominated by the warp structure in the NW disk.
The the Gaussian fit center velocity of H$\alpha$ shows an offset by +24.2 km s$^{-1}$ 
from the systemic velocity obtained from the HI emission.
We speculate that a minor/mini-merger event may exist in the NW disk of AGC 102004.

(2) We determined the gas-phase metallicity, 12+log(O/H), using [{\ion{N}{II}}]$\lambda$6583/H$\alpha$ diagnostic.
The mean value of 12+log(O/H) is 8.30 $\pm$ 0.19 over the whole galaxy.
The metallicity in the outer disk is lower than the detection limit of 7.72.
The metallicity gradient in AGC 102004 may signify the occurrence 
of a merger or metal-poor gas accretion event.

(3) CO(2-1) has non-detection in AGC 102004.
The upper limit of molecular gas mass in AGC 102004 is 2.1 $\times$ 10$^7$ M$\odot$.
Variations in the chosen X$_{\rm CO}$ value do not significantly change our results. 
The M$_{\rm H_2}$/M$^{\rm corr}_{\rm HI}$ of AGC 102004 is lower than that of normal galaxies.
The molecular gas depletion time of AGC 102004 is shorter than that of the nearby disk galaxies. 
The inefficiency of star formation in AGC 102004 is likely caused by the low efficiency in converting molecules from atomic gas.
We compare CO luminosities of LSBGs and other normal galaxies.
CO luminosities in LSBGs are lower than that of normal galaxies in the relationship between the CO luminosity and M$^{\rm corr}_{\rm HI}$.
Low metallicity environments may contribute to the low detection rate of CO in LSBGs.

\section*{Acknowledgments}
We thank the anonymous referee for a number of very
constructive comments.
We thank Dr. Don Neill, Prof. Min Du, and Dr. Xiao Cao for their 
insightful comments and/or useful communications during the preparation of the manuscript.
We thank Prof. E. M. Di Teodoro help for in resolving the problem we met using $^{3D}$BAROLO software.

J.W. acknowledges National Key R\&D Program of China (Grant No. 2023YFA1607904) and the National
Natural Science Foundation of China (NSFC) grants 12033004, 12333002, 12221003.
T.\,C. acknowledges the China Postdoctoral Science Foundation (Grant No. 2023M742929),
and the NSFC grants 12173045 and 12073051.
G.\,G. acknowledges the ANID BASAL projects ACE210002 and FB210003. 
C.\,C. is supported by the NSFC, No. 11803044, 11933003, 12173045 and acknowledge the science research grants 
from the China Manned Space Project with NO. CMS-CSST-2021-A05.
H.\,W. acknowledges the NSFC grant No.\,11733006 and 12090041.
This work is sponsored (in part) by the Chinese Academy of Sciences (CAS), 
through a grant to the CAS South America Center for Astronomy (CASSACA).
This work is supported (in part) by NSFC grants 12090040 and 12090041,
and the Strategic Priority Research Program of the Chinese Academy of Sciences, Grant No. XDB0550100.

This research uses data obtained through the Telescope Access Program (TAP), 
which has been funded by the TAP association, including the Center for Astronomical Mega-Science CAS (CAMS), 
XMU, PKU, THU, USTC, NJU, YNU, and SYSU.

The James Clerk Maxwell Telescope (JCMT) is operated by the East Asian 
Observatory on behalf of The National Astronomical Observatory of Japan; 
Academia Sinica Institute of Astronomy and Astrophysics; 
the Korea Astronomy and Space Science Institute; 
the National Astronomical Research Institute of Thailand; 
Center for Astronomical Mega-Science 
(as well as the National Key R\&D Program of China with No. 2017YFA0402700).
 Additional funding support is provided by 
 the Science and Technology Facilities Council of the United Kingdom and 
 participating universities and organizations in the United Kingdom and Canada.
 N$\bar{\rm a}$makanui was constructed and funded by ASIAA in Taiwan, with funding for the 
 mixers provided by ASIAA and at 230GHz by EAO. 
 The N$\bar{\rm a}$makanui instrument is a backup receiver for the GLT.

\bibliographystyle{apj}
\bibstyle{thesisstyle}
\bibliography{main}

\end{document}